\documentclass[12pt, english, dvipsnames, table]{wlpeerj} 
\usepackage{xcolor}
\definecolor{vlg}{rgb}{0.88, 0.88, 0.88}
\usepackage{palatino}
\usepackage{subfiles}
\usepackage[english]{babel} 
\usepackage[comma]{natbib}
\usepackage{url}
\usepackage{xspace}
\usepackage{amsmath}
\usepackage{float}
\usepackage{verbatim}
\usepackage{fancyvrb}
\usepackage{tabularx}
\usepackage{pdflscape}
\usepackage{afterpage}
\usepackage[colorlinks=true, citecolor = purple, linkcolor = purple]{hyperref}
\usepackage{environ}
\usepackage{marginnote}
\usepackage{lineno}
\usepackage{setspace}
\usepackage{framed}

\reversemarginpar
\raggedright
\usepackage{indentfirst}
\newcommand{\pval}{p-value\xspace}
\newcommand{\pvals}{p-values\xspace}

\title{I can see clearly now: reinterpreting statistical significance}

\author[1]{Jonathan Dushoff}
\author[1]{Morgan P. Kain}
\author[1,2]{Benjamin M. Bolker}

\affil[1]{Department of Biology, McMaster University, 1280 Main Street West, Hamilton, Ontario L8S 4K1 Canada}
\affil[2]{Department of Mathematics and Statistics, McMaster University, 1280 Main Street West, Hamilton, Ontario 
L8S 4L8 Canada}

\corrauthor[1]{Jonathan Dushoff}{dushoff@mcmaster.ca}

\begin{document}

\flushbottom
\maketitle


\doublespacing

\subsection*{Abstract}

\noindent Null hypothesis significance testing remains popular despite decades of concern about misuse and misinterpretation. We believe that much of the problem is due to language: significance testing has little to do with other meanings of the word ``significance''. Despite the limitations of null-hypothesis tests, we argue here that they remain useful in many contexts as a guide to whether a certain effect can be seen \emph{clearly} in that context (e.g. whether we can clearly see that a correlation or between-group difference is positive or negative). We therefore suggest that researchers describe the conclusions of null-hypothesis tests in terms of statistical ``clarity'' rather than statistical ``significance''. This simple semantic change could substantially enhance clarity in statistical communication.

\subsection*{Key Words}

\noindent Statistical philosophy; Statistical clarity; Hypothesis testing; \pval

\clearpage

\subsection*{Introduction}

\noindent Statisticians and scientists have bemoaned the shortcomings of null hypothesis significance testing (NHST) for nearly a century \citep{Cohen1994}. Books and articles proposing the de-emphasis or abandonment of the \pval have been cited thousands of times \citep{Cohen1994, Goodman1999, Wilkinson1999, ZiliakandMcCloskey2008, WassersteinandLazar2016}. These works plead for a focus on effect sizes and confidence intervals, and point out that null effects that truly have zero magnitude are unrealistic or impossible in most fields outside of the hard physical sciences \citep{Meehl1990, Tukey1991, Cohen1994}. Yet, \pvals without confidence intervals (or even effect sizes) and references to null effects still pervade the scientific literature at all levels up to and including articles in high-impact journals.

In a meta-analysis of 356 studies \citet{Bernardietal.2017} found that 72\% of studies contained an ambiguous use of the term ``significant'', 49\% interpreted non-significant effects as zero effects, and 44\% failed to report a comprehensible effect size. The misuse and misinterpretation of NHST is so frequent that there have been recent calls for drastically reducing \citep{SzucsandIoannidis2017} or abandoning \citep{McShaneetal.2017} its use. Other prescriptions have included the complete abandonment of frequentist statistics \citep{The2011}, or the use of a stricter significance threshold (e.g. $p < 0.005$: \cite{Benjaminetal.2018}); however, the former seems impractical, while the latter is unlikely to reduce the misuse and misinterpretation of \pvals, or the publication bias imposed by any \pval threshold \citep{Ridleyetal.2007}.

Here, we argue that NHST remains useful, and that pervasive misuse can be reduced through a linguistic change: using the language of statistical ``clarity'' instead of statistical ``significance''.

\subsection*{The null hypothesis is false}

\noindent In most biological studies, the null hypothesis is known or believed to not be strictly true. Even in cases where the null hypothesis is sensible (e.g., particle physics, \cite{Staley2017}), NHST does not provide evidence that a difference is exactly zero. This being the case, it is worth asking how NHST has survived ``if it is as idiotic as \ldots long believed'' \citet[cited in \cite{Kramer2011}]{ZiliakandMcCloskey2008}.

The value of NHST can be seen in something like a permutation-based t-test \cite[Chapter 1]{GoodBook}: it provides a simple, robust framework to ask whether we can tell \emph{which} mean is bigger. More generally, testing the null hypothesis is a proxy for asking whether we clearly see a signal of how our data differs from it. In many cases, this comes down simply to whether we can be confident of the \emph{sign} of a difference or a correlation coefficient \citep{robinson2001past}. In other cases (e.g., a one-way ANOVA), it may not be simple to describe the difference, but NHST is still a reasonable, accepted way to evaluate whether an effect has been seen clearly.

The ``idiocy'', if any, comes in the interpretive step. A statistical fact (``we have \emph{seen} a difference between the groups'', which should immediately prompt the question ``what have you learned \emph{about} that difference?'') is interpreted as a scientific fact (``there \emph{is} a `significant' difference between the groups), which is often seen as an end in itself: ``we showed that the groups differ''.

\subsection*{The \pval is a property of the study}

\noindent We often see sentences like, ``X et al. showed that there is no significant effect of Y on Z'' with the implication that this effect can now be assumed to be absent (or unimportant). In fact, the sentence is erroneous even before we get to the implication: significance tests provide information about \emph{a data set} -- that is, about a study, not about the study system \citep{HoenigandHeisey2001}. Indeed, a very small effect can lead to $p < 0.05$, when data is abundant (or noise is small); or a very large one can lead to $p > 0.05$ when the sample is small or noisy. 

The statement ``X et al. showed that Y has a statistically significant effect on Z'' is similarly misleading. Frequentist statistics effectively assume that the effect is present (or at least, admit that it can't be disproven). The question is whether it is seen in a particular data set. The statement ``X et al. were able to see the effect of Y on Z'' is not only more accurate, but it appropriately implies that something is missing: \emph{What} effect did they see?

\subsection*{Statistical clarity}

\noindent The language of ``statistical clarity'' could help researchers escape various logical traps while interpreting the results of NHST, allowing for the continued use of NHST as a simple, robust method of evaluating whether a data signal is clear (see \cite{Abelson1997} for arguments for NHST). The use of ``significance'' to describe the results of hypothesis tests is deeply, and sometimes subtly, misleading, because it is at odds with other meanings of the word: the \pval is not an accurate gauge of whether a result is large in magnitude, biologically important, or relevant. ``Clarity,'' on the other hand, is an apt term for what NHST actually evaluates. \citet{jones2000sensible} and \citet{robinson2001past} suggest that researchers should report $p > 0.05$ using language such as ``the direction of the differences among the treatments was undetermined''. This is a step in the right direction. Replacing ``significance'' with ``clarity'' takes this idea further, and has the promise to substantially improve statistical communication.

For example, the sentence ``X et al. showed that the effect of Y on Z is statistically unclear'', is noticeably awkward. It seems less like a statement about the study system, and suggests the more straightforward ``did not find a statistically clear effect.'' Similarly, ``We did not find a clear difference in response between the control and sham groups'' is both more colloquial and harder to transform into a misleading statement than ``We did not find a significant difference \ldots''. \citet{Bernardietal.2017} complained that ``\ldots sociological and social significance are sacrificed on the altar of statistical significance''. Describing statistical tests in terms of clarity would allow ``significant'' to reclaim its common English definition and reduce conflation between statistical results and substantive significance.

Descriptions of statistical results using the language of clarity should begin with reference to the effect. For example, ``The difference between the control and treatment group was not statistically clear.'' Table~\ref{quotetab} shows published examples of statements that misinterpret \pvals in three different ways and demonstrates how to rephrase them in the language of clarity.

\clearpage

\newcommand{\ourcomment}[1]{[\emph{#1}]}
\begin{table}
\setlength\tabcolsep{1cm}.
\begin{tabular}{p{7.0cm}p{7.0cm}}
\textbf{Language from published articles} & \textbf{Rewritten using ``clarity''} \\
\hline\\

\multicolumn{2}{c}{\emph{\textbf{Accepting the null hypothesis ($\boldsymbol{p > 0.05 \nRightarrow}$ no effect)}}} \\
\hline

Toxins accumulate after acute exposure but have no effect on behaviour
& Toxins accumulate after acute exposure but their effects on behaviour are statistically unclear
\\

\rowcolor{vlg}
There was no effect of elevated carbon dioxide on reproductive behaviors 
& The effect of elevated carbon dioxide on reproductive behaviors was statistically unclear
\\

The finding that species richness showed no significant relationship with the area of available habitat is surprising because richness is usually strongly influenced by landscape context 
& Although species richness is usually strongly influenced by landscape context, we were unable to find a statistically clear relationship in this study
\\ \\

\multicolumn{2}{c}{\emph{\textbf{Inferring weak effects from large \pvals}} \citep{WassersteinandLazar2016}}
\\
\hline
\ldots\ differences between treatment and control groups were nonsignificant, with P values of at least 0.3, and most in the range $0.7 \leq P \leq 0.9$.
& \ldots\ differences between treatment and control groups were not statistically clear (all $P > 0.05$) \ourcomment{confidence intervals would also be valuable here!}
\\ \\

\multicolumn{2}{c}{\emph{\textbf{The difference between ``clear'' and ``not clear'' is not clear}} \citep{GelmanandStern2006}}
\\
\hline

This correlation was significant in males ($\rho=0.35$, P \textless 0.05) but not females ($\rho=0.35$, NS). \ldots \ourcomment{The authors later write as though they have demonstrated a difference between males and females}
& Although males and females show the same correlation coefficient ($\rho=0.35$), the sign of the coefficient is statistically clear only in males  \ldots \ourcomment{This phrasing may suggest to the authors that confidence intervals are called for.}
\\
\rowcolor{vlg}
\ldots risk of low BMD [bone mineral density] remained greater in HCV-coinfected women versus women with HIV alone
(adjusted OR 2.99, 95\% CI 1.33–6.74), but no association was found between HCV coinfection and low BMD in men 
(adjusted OR 1.26, 95\% CI 0.75–2.10). \ldots 
The precise mechanisms for the association between viral hepatitis and low BMD in HIV-infected women but not men remain unclear.
& 
\ldots  risk of low BMD [bone mineral density] remained greater in HCV-coinfected women versus women with HIV alone
(adjusted OR 2.99, 95\% CI 1.33–6.74), but the association between HCV coinfection and low BMD in men was not
statistically clear (adjusted OR 1.26, 95\% CI 0.75–2.10). \ldots  Pursuing biological differences between women and men in the effect of HIV on BMD would be premature given these results.
\\
\end{tabular}
\captionof{table}{Examples of misleading language in peer-reviewed papers (citations available by request), and revisions using our proposed  language of clarity.}
\label{quotetab}
\end{table}

\clearpage

\subsection*{Conclusions}

\noindent We believe that NHST is useful as a simple, robust way to ask whether an effect can be seen clearly in a particular data set \citep{robinson2001past}, and that careful, clarity-based language can reduce misinterpretation and miscommunication.

We agree with \citet{Cohen1994} and others \citep{Goodman1999, ZiliakandMcCloskey2008, WassersteinandLazar2016}, that scientific communication and understanding will be improved by a shift away from \pvals to effect sizes and confidence intervals. We argue that the use of ``statistical clarity'' reinforces the need for confidence intervals and effect sizes by making it clearer that bald statements about \pvals are insufficient. The statement ``The difference between our control and treatment groups was not statistically clear ($p = 0.30$)'' is noticeably incomplete; an effect size and confidence interval are required to complete the story.

Improving language will not by itself solve all of the known problems with current statistical practice. We echo previous statements in favor of ``neglected factors'' (prior and related evidence, plausibility of mechanism, study design and data quality, real world benefits, novelty and other factors) \citep{McShaneetal.2017} and reporting of \emph{a priori} analysis of statistical power to avoid emphasis on implausibly large effects given low statistical power \citep[the ``winner's curse''][]{GelmanandCarlin2014, SzucsandIoannidis2017, Bernardietal.2017}. Additionally, we support the idea of writing a statistical journal that chronicles all steps in the analytical process \citep{Kassetal.2016}, and clearly delineating the boundary between inferences based on \emph{a priori} hypotheses and discoveries from \emph{post hoc} data exploration.

Whether or not our recommendations are broadly adopted by authors, reviewers, and editors, they can be useful for individual researchers who want to help themselves think clearly about NHST results. We have found that rephrasing NHST statements that we encounter (in the literature, or in seminar presentations) in terms of clarity has already helped us with both interpretation and communication.

\subsection*{Acknowledgments}

\noindent We thank members of the Dushoff and Bolker labs for helpful comments on the first draft of the manuscript.

\clearpage

\bibliography{stat_signif}


\end{document}